\documentclass{article}
\usepackage{epsfig}
\usepackage{graphicx}
\usepackage{graphics}
\usepackage{amsfonts}
\usepackage{marvosym}
\usepackage{pstricks}

\DeclareFontFamily{OT1}{pzc}{}
\DeclareFontShape{OT1}{pzc}{m}{it} {<-> s * [1.2] pzcmi7t}{}
\DeclareMathAlphabet{\mathpzc}{OT1}{pzc}{m}{it}
\newcommand{\db}{\mathpzc{d}_\mathcal{B}}

\begin{document}
\title{Methods of Hierarchical Clustering}
\author{Fionn Murtagh (1, 2) and Pedro Contreras (2)\\
(1) Science Foundation Ireland, Wilton Place, Dublin 2, Ireland \\
(2) Department of Computer Science, Royal Holloway, University of 
London \\ Egham TW20 0EX, England \\
Email: fmurtagh@acm.org}

\maketitle

\begin{abstract}
We survey agglomerative 
hierarchical clustering algorithms and discuss 
 efficient implementations that are available in R and other 
software environments.  We look at hierarchical self-organizing 
maps, and mixture models.  We review grid-based clustering, focusing
on hierarchical density-based approaches.  Finally we describe 
a recently developed very efficient (linear time) hierarchical 
clustering algorithm, which can also be viewed as a hierarchical 
grid-based algorithm.  
\end{abstract}

\section{Introduction}

Agglomerative hierarchical clustering has been the dominant 
approach to constructing embedded classification schemes.  
It is our aim to 
direct the reader's attention to practical
algorithms and methods --
both efficient (from the computational and storage points of view)
and effective (from the application point of view).  
It is often helpful to distinguish between {\em method}, involving a
compactness criterion and the target structure of a 2-way tree 
representing the partial order on subsets of the power set; as
opposed to an {\em implementation}, which relates to the detail
of the algorithm used.  

As with many 
other multivariate techniques, the objects to be classified have
numerical measurements on a set of variables or attributes.
Hence, the  analysis is carried out on the rows of an array or matrix.
If we do not have a matrix of numerical values to begin with, then
it may be necessary to skilfully construct such a matrix.
The objects, or rows of the matrix, can be viewed as vectors in
a multidimensional space (the dimensionality of this space being
the number of variables or columns).  A geometric framework of
this type is not the only one which can be used to formulate
clustering algorithms.  
Suitable alternative forms of storage of a rectangular array of
values are not inconsistent with viewing the problem in geometric
terms (and in matrix terms -- for example, expressing the 
adjacency relations in a graph). 

Motivation for clustering in general, covering hierarchical clustering
and applications, includes the following:
analysis of data; interactive user interfaces; storage and retrieval;
and pattern recognition.

Surveys of clustering with coverage also of hierarchical clustering 
include  Gordon (1981), March (1983), Jain and Dubes (1988), Gordon (1987), 
Mirkin (1996), 
Jain, Murty and Flynn (1999), and Xu and Wunsch (2005).   
Lerman (1981) and Janowitz (2010) present overarching 
reviews of clustering including through use of lattices that generalize
trees.  
The case for the central role of hierarchical clustering in information 
retrieval 
was made by van Rijsbergen (1979) and continued in the work of Willett 
(e.g.\ Griffiths et al., 1984)
and others.  Various mathematical views of hierarchy, all expressing 
symmetry in one way or another, are explored in Murtagh (2009). 

This article is organized as follows.  

In section \ref{sect5} we look at the issue of normalization of 
data, prior to inducing a hierarchy on the data.

In section \ref{sect1} some historical remarks and motivation are provided 
for hierarchical agglomerative clustering.

In section \ref{sect3}, we 
discuss the Lance-Williams formulation of a wide range of algorithms,
and how these algorithms can be expressed in graph theoretic 
terms and in geometric terms.  In section \ref{sect2}, we describe
the principles of the reciprocal nearest neighbor and nearest neighbor
chain algorithm, to support building a hierarchical clustering 
in a more efficient manner compared to the Lance-Williams or general 
geometric approaches.  

In section \ref{sect6} we overview the hierarchical Kohonen self-organizing
feature map, and also hierarchical model-based clustering.  We conclude 
this section with some reflections on divisive hierarchical clustering, 
in general.  

Section \ref{sect7} surveys developments in grid- and density-based
clustering.  The following section, section \ref{sect8}, 
presents a recent algorithm of this type, which is particularly
suitable for the hierarchical clustering of massive data
sets.  

\section{Data Entry: Distance, Similarity and Their Use}
\label{sect5}

Before clustering comes the phase of data measurement, or measurement of the
observables.  Let us look at some important 
considerations to be taken into account.  These considerations 
relate to the metric or other spatial embedding, comprising 
the first phase of the data analysis {\em stricto sensu}.  

To group data we need a way to measure the elements and their 
distances relative 
to each other in order to decide which elements belong to a group. 
This can be a 
similarity, although on many occasions a dissimilarity 
measurement, or a ``stronger'' distance, is  used. 

A distance between any pair of vectors or points $i, j, k$ 
satisfies the properties of: 
symmetry, $d(i,j) = d(j,k)$; positive definiteness, $d(i,j) 
> 0$ and $d(i,j) = 0 $ iff $ i = j$; and the triangular 
inequality, $d(i,j) \leq d(i,k) + d(k,j)$.  If the triangular
inequality is not taken into account, we have a dissimilarity.
Finally a similarity is given by $s(i,j) = $  
max$_{i,j} \{ d(i,j) \} - 
d(i,j)$.  

When working in a vector space a traditional way to measure 
distances is a Minkowski distance, which is a family of metrics 
 defined as follows:

\begin{equation}
	L_{p}(\mathbf{x}_{a}, \mathbf{x}_{b}) = (\sum_{i=1}^{n} |\mathbf{x}_{i,a} - 
	\mathbf{x}_{i,b}|^{p})^{1/p};\;\forall~p \geq 1,~ p \in \mathbb{Z^+},
\end{equation}
where $\mathbb{Z^+}$ is the set of positive integers.

The Manhattan, Euclidean and Chebyshev distances (the latter is 
also called maximum distance) 
are special cases of the Minkowski distance when 
$p=1,~p=2$ and $p \to \infty$.  

As an example of similarity we have 
the {\em cosine} similarity, which gives the 
angle between two vectors. This is widely used in text retrieval to match 
vector queries to the dataset. The smaller the angle between a query vector 
and a document vector, the closer a query is to a document. The 
normalized cosine 
similarity is defined as follows:

\begin{equation}
	s(\mathbf{x}_{a}, \mathbf{x}_{b}) = \cos(\theta) = \frac{\mathbf{x}_{a} 
	\cdot \mathbf{x}_{b}} {\|\mathbf{x}_{a}\|\|\mathbf{x}_{b}\|}
\end{equation}	
where $\mathbf{x}_{a} \cdot \mathbf{x}_{b}$ is the dot product and $\|\cdot\|$ 
the norm. 

Other relevant distances are the Hellinger, variational, Mahalanobis and 
Hamming distances. Anderberg (1973)
gives a good review of measurement and metrics, 
where their interrelationships are also discussed. Also
Deza and Deza (2009) have produced a comprehensive list of distances in their 
{\em Encyclopedia of Distances}. 

By mapping our input data into a Euclidean space, where each object is 
equiweighted, we can use a Euclidean distance for the clustering that 
follows.  Correspondence analysis is very versatile in determining a Euclidean,
factor space from a wide range of input data types, including frequency
counts, mixed qualitative and quantitative data values, ranks or scores,
and others.  
Further reading on this 
is to be found in Benz\'ecri (1979), Le Roux and Rouanet
(2004) and Murtagh (2005).

\section{Agglomerative Hierarchical Clustering Algorithms: Motivation}
\label{sect1}

Agglomerative hierarchical clustering 
algorithms can be characterized as {\em greedy}, in the algorithmic 
sense.  
A sequence of irreversible algorithm steps 
is used to construct the desired data structure.  Assume that a pair of 
clusters, including possibly singletons, is merged or agglomerated at 
each step of the algorithm.  Then the following are equivalent views of 
the same output structure constructed on $n$ objects: a set of $n-1$ 
partitions, starting with the fine partition consisting of $n$ classes 
and ending with the trivial 
partition consisting of just one class, the entire 
object set; a binary tree 
(one or two child nodes at each non-terminal node) commonly referred to 
as a dendrogram; a partially ordered
set (poset) which is a subset of the power set of the $n$ objects; and 
an ultrametric topology on the $n$ objects. 

An ultrametric, or tree metric, defines a stronger topology compared 
to, for example, a Euclidean metric geometry.   For three points, $i, j, k$,
metric and ultrametric respect the properties of symmetry 
($d$, $d(i,j) = d(j,i)$) and positive definiteness ($d(i,j) > 0$ and 
if $d(i,j) = 0$ then $i = j$).   A metric though 
(as noted in section \ref{sect5})
satisfies the triangular
inequality, $d(i,j) \leq d(i,k) + d(k,j)$ while an ultrametric satisfies
the strong triangular or ultrametric (or non-Archimedean), inequality, 
$d(i,j) \leq  \mbox{max} \{ d(i,k), d(k,j) \}$.  In section \ref{sect5},
above, there was further discussion on metrics.  

%

The single linkage hierarchical clustering approach outputs a set of
clusters (to use graph theoretic terminology, a set of maximal connected
subgraphs) at each level -- or for each threshold value which produces
a new partition.  The single linkage method with
which we begin is one of the oldest methods, its origins being traced to 
Polish researchers in the 1950s (Graham and Hell, 1985).  
The name {\em single linkage} arises since the interconnecting
dissimilarity between two clusters or components is defined
as the least interconnecting dissimilarity between a member of one and 
a member of the other.  Other hierarchical clustering methods are 
characterized by other functions of the interconnecting linkage
dissimilarities.


As early as the 1970s, it was held that about $75\%$ of all 
published work on clustering employed hierarchical
algorithms (Blashfield and Aldenderfer, 1978). Interpretation
of the information contained in a dendrogram is often of one
or more of the following kinds: set inclusion relationships,
partition of the object-sets, and significant clusters.

Much early work on hierarchical clustering was in the field of 
biological taxonomy, from the 1950s and more so from the 1960s
onwards.  The central reference in this area, the first edition
of which dates from the early 1960s, is Sneath and 
Sokal (1973).  One major interpretation of hierarchies has been
the evolution relationships between the organisms under study.  It
is hoped, in this context, that a dendrogram provides a sufficiently
accurate model of underlying evolutionary progression. 

A common interpretation made of hierarchical clustering is to
derive a partition. A further type of interpretation is instead to 
detect maximal (i.e.\ disjoint) clusters of
interest at varying levels of the hierarchy.  Such an approach is
used by Rapoport and Fillenbaum (1972) in a clustering of colors
based on semantic attributes.  Lerman (1981) developed an approach 
for finding significant clusters at varying levels of a hierarchy,
which has been widely applied.  By developing a wavelet transform 
{\em on} a dendrogram (Murtagh, 2007), 
which amounts to a wavelet trasform in the
associated ultrametric topological space, the most important -- in 
the sense of best approximating -- clusters can be determined. 
Such an approach is a topological one (i.e., based on sets and their 
properties) as contrasted with more widely used optimization or 
statistical approaches.  

In summary, a dendrogram collects together many of the proximity
and classificatory relationships in a body of data.  It is a convenient 
representation which answers such questions as: ``How many useful 
groups are 
in this data?'', ``What are the salient interrelationships present?''.  
But it can be noted that differing answers can feasibly be
provided by a dendrogram for most of these questions, depending on the
application.  

\section{Agglomerative Hierarchical Clustering Algorithms}
\label{sect3}

A wide range of agglomerative hierarchical clustering 
algorithms have been proposed at one time or another.
Such hierarchical algorithms may be conveniently broken down into
two groups of methods.  The first group is that of
linkage methods --  the single, complete, weighted
and unweighted average linkage methods. These are methods for which a
graph representation can be used.  Sneath and Sokal (1973) may be 
consulted for many other graph representations of the stages in
the construction of hierarchical clusterings.  

The second group of hierarchical clustering methods are methods 
which allow the cluster centers to be specified (as an average 
or a weighted average of the member vectors of the cluster).  
These methods include the centroid, median and minimum variance methods.

The latter may be specified either in terms of dissimilarities, alone,
or alternatively in terms of cluster center coordinates and dissimilarities.
A very convenient formulation, in dissimilarity terms, which 
embraces all the hierarchical methods mentioned so far, is the
{\em Lance-Williams dissimilarity update formula}.  If points (objects)
$i$ and $j$ are agglomerated into cluster $i \cup j$, then we must simply
specify the new dissimilarity between the cluster and all other
points (objects or clusters).  The formula is:

$$        d(i \cup j,k)  =  \alpha_i d(i,k) + \alpha_j d(j,k) + \beta
                d(i,j) +  \gamma \mid d(i,k)-d(j,k) \mid $$ where 
$\alpha_i$, $\alpha_j$, $\beta$, and $\gamma$ define the agglomerative 
criterion.  Values
of these are listed in the second column of Table \ref{tabhier}.
\begin{table}
\begin{tabular}{|l|lll|} \hline
Hierarchical & Lance and Williams & Coordinates    & Dissimilarity \\
clustering   & dissimilarity      & of center of   & between cluster \\
methods (and & update formula    & cluster, which & centers $g_i$ and $g_j$ \\
aliases)    &                    &   agglomerates &         \\
             &                    &   clusters $i$ and $j$ &  \\  \hline  
Single link  &  $\alpha_i = 0.5$  &                & \\
(nearest     &  $\beta = 0$       &                & \\  
neighbor)  &  $\gamma = -0.5$   &                & \\
             &  (More simply:     &                & \\
             &  $min\{d_{ik},d_{jk}\}$) &          & \\   \hline
Complete link & $\alpha_i = 0.5$  &                & \\
(diameter)  &  $\beta = 0$       &                & \\
             &  $\gamma = 0.5$    &                & \\
             &  (More simply:     &                & \\
             &  $max\{d_{ik},d_{jk}\}$) &          & \\   \hline
Group average & $\alpha_i = {{\mid i \mid} \over {\mid i \mid + \mid j \mid}}$
                                  &                & \\
(average link, & $\beta = 0$      &                & \\
UPGMA)      &  $\gamma = 0$      &                & \\    \hline
McQuitty's   &  $\alpha_i = 0.5$  &                & \\
method       &  $\beta = 0$       &                & \\
(WPGMA)     &  $\gamma = 0$      &                & \\    \hline
Median method & $\alpha_i = 0.5$  & ${\bf g} = {{{\bf g}_i + {\bf g}_j}
                                             \over {2} }$
                                   & $\| {\bf g}_i - {\bf g}_j \|^2$ \\
(Gower's,    &  $\beta = -0.25$    &               & \\
WPGMC)      &  $\gamma = 0$      &                & \\  \hline
Centroid     &  $\alpha_i = { {\mid i \mid} \over {\mid i \mid + \mid j \mid}}$
              & $ {\bf g} ={{ \mid i \mid {\bf g}_i + \mid j \mid {\bf g}_j}
               \over { \mid i \mid + \mid j \mid }}$
               &  $ \| {\bf g}_i - {\bf g}_j \|^2 $  \\
(UPGMC)     &  $\beta = - {{\mid i \mid \mid j \mid} \over {(\mid i \mid + 
                                                   \mid j \mid)^2}} $ &    & \\
             &  $\gamma = 0$       &                & \\   \hline
Ward's method & $\alpha_i = { {\mid i \mid + \mid k \mid } \over 
                              {\mid i \mid + \mid j \mid + \mid k \mid }}$ 
              & $ {\bf g} ={{ \mid i \mid {\bf g}_i + \mid j \mid {\bf g}_j}
               \over { \mid i \mid + \mid j \mid }}$
               &  $  {{ \mid i \mid \mid j \mid } \over { \mid i \mid + 
                   \mid j \mid }} \| {\bf g}_i - {\bf g}_j \|^2 $   \\
(minimum var- &  $\beta = - { {\mid k \mid} \over {\mid i \mid + \mid j \mid 
                               + \mid k \mid}}$
                                   &                 &  \\
iance, error  &  $\gamma = 0$      &                 &  \\
sum of squares) &                   &                &  \\  \hline
\end{tabular}
Notes: $\mid i \mid$ is the number of objects in cluster $i$.
${\bf g}_i$ is a vector in $m$-space ($m$ is the set of
attributes), -- either an intial point or a cluster center.
$\|.\|$ is the norm in the Euclidean metric.  The names UPGMA, 
etc.\ are due to Sneath and Sokal (1973).  Coefficient $\alpha_j$,
with index $j$, is defined identically to coefficient $\alpha_i$
with index $i$.  Finally, the Lance
and Williams recurrence formula is (with $\mid . \mid$ 
expressing absolute value):
$$ d_{i \cup j,k} = \alpha_i d_{ik} + \alpha_j d_{jk}
  + \beta d_{ij} + \gamma \mid d_{ik} - d_{jk} \mid . $$
  \caption{Specifications of seven hierarchical clustering methods.}
\label{tabhier}
\end{table}
In the case
of the single link method, using $\alpha_i = \alpha_j = {1 \over 2}$, $\beta = 
0$, and 
$\gamma = -{1 \over 2}$ gives us

$$       d(i \cup j,k)  =  {1 \over 2} d(i,k) + {1 \over 2} d(j,k) - 
             {1 \over 2} \mid 
                d(i,k)-d(j,k)\mid $$ which,
it may be verified, can be rewritten as

$$       d(i \cup j,k)  =  {\rm min} \ \{ d(i,k),d(j,k) \}.$$ 

Using other update
formulas, as given in column 2 of Table \ref{tabhier}, allows the other
agglomerative methods to be implemented in a very similar way
to the implementation of the single link method.

In the case of the methods which use cluster centers, we have
the center coordinates (in column 3 of Table \ref{tabhier}) and 
dissimilarities as defined between cluster centers (column 4 of
Table \ref{tabhier}).  The Euclidean distance must be used
for equivalence between the two approaches.  In the case of
the {\em median method}, for instance, we have the following (cf.\
Table \ref{tabhier}).

Let ${\bf a}$ and ${\bf b}$ be two points (i.e.\ $m$-dimensional vectors: 
these are objects
or cluster centers) which have been agglomerated, and let ${\bf c}$ be another
point.  From the Lance-Williams dissimilarity update formula, using
squared Euclidean distances, we have:

 \begin{equation}
    \begin{array}{l@{\;\;\;=\;\;\;}l}   
d^2(a \cup b,c)  & 
{{d^2(a,c)}\over{2}} + {{d^2(b,c)}\over{2}} - {{d^2(a,b)}\over{4}}  \\[5pt]
  &
{{ \| {\bf a}-{\bf c} \|^2} \over {2}} + {{ \|{\bf b}-{\bf c}\|^2} \over {2}}
         -  
        {{ \|{\bf a}-{\bf b}\|^2} \over {4}}   .
     \end{array}
    \end{equation}

The new cluster center is $({\bf a}+{\bf b})/2$, so that its distance 
to point $\bf c$ is

\begin{equation}
       \|{\bf c} - {{ {\bf a}+{\bf b}} \over 2}\| ^2  . 
\end{equation}

That these two expressions are identical is readily verified.  The 
correspondence between these two perspectives on the one agglomerative
criterion is similarly proved for the centroid and minimum variance
methods.

For cluster center methods, and with suitable alterations for graph methods, 
the following algorithm is an alternative to the general dissimilarity based 
algorithm.  The latter may be described as a ``stored dissimilarities 
approach'' (Anderberg, 1973).

\vspace{.25in}

{\bf Stored data approach}
\begin{description}

\item[Step 1] Examine all interpoint dissimilarities, and form cluster from
        two closest points.

\item[Step 2] Replace two points clustered by representative point (center
        of gravity) or by cluster fragment.

\item[Step 3] Return to step 1, treating clusters as well as remaining
        objects, until all objects are in one cluster.
\end{description}
\vspace{.25in}

In steps 1 and 2, ``point'' refers either to objects or clusters, both
of which are defined as vectors in the case of cluster center methods.
This algorithm is justified by storage considerations, since we
have $O(n)$ storage required for $n$ initial objects and $O(n)$ storage
for the $n-1$ (at most) clusters.  In the case of linkage methods, the
term ``fragment'' in step 2 refers (in the terminology of graph theory)
to a connected component in the case
of the single link method and to a clique or complete subgraph in 
the case of the complete link method.  Without consideration of 
any special algorithmic ``speed-ups'', the overall complexity of the
above algorithm is $O(n^3)$ due to the repeated calculation of dissimilarities
in step 1, coupled with $O(n)$ iterations through steps 1, 2 and 3.
While the stored data algorithm is instructive, it does not lend itself
to efficient implementations.  In the section to follow, we 
look at the reciprocal nearest 
neighbor and mutual nearest neighbor algorithms which can be 
used in practice for implementing agglomerative hierarchical clustering
algorithms.  

Before concluding this overview of agglomerative hierarchical 
clustering algorithms, we will describe briefly the minimum variance method.  

The variance or spread of a set of points (i.e.\ the average of the 
sum of squared 
distances from the center) has been a point of departure for
specifying clustering algorithms.  Many of these algorithms, --
iterative, optimization algorithms as well as the hierarchical,
agglomerative algorithms -- are described and appraised in
Wishart (1969).  The use of variance in a clustering criterion
links the resulting clustering to other data-analytic techniques
which involve a decomposition of variance, and make   
the minimum variance agglomerative
strategy particularly suitable for synoptic clustering.  
Hierarchies are also more balanced with this agglomerative 
criterion, which is often of practical advantage. 

The minimum variance method produces clusters which satisfy compactness
and isolation criteria.  These criteria are incorporated into the
dissimilarity.
We seek to agglomerate two clusters, $c_1$ and $c_2$, into
cluster $c$ such that the within-class variance of the partition thereby
obtained is minimum.  Alternatively, the between-class variance of
the partition obtained is to be maximized. Let
$P$ and $Q$ be the partitions prior to, and subsequent to, the agglomeration;
let $p_1$, $p_2$, \dots be classes of the partitions:

\begin{displaymath}
\begin{array}{l@{\;\;=\;\;}l}
    P  &  \{ p_1, p_2, \dots , p_k, c_1, c_2 \} \\
    Q  &  \{ p_1, p_2, \dots , p_k, c \} .
\end{array}
\end{displaymath}

Letting $V$ denote {\em variance}, then 
in agglomerating two classes of $P$, the variance of the resulting
partition (i.e.\ $V(Q)$ ) will necessarily decrease: therefore in
seeking to minimize this decrease, we simultaneously achieve a 
partition with maximum between-class variance.  The criterion to
be optimized can then be shown to be:
\begin{displaymath}
 \begin{array}{l@{\;\;\;=\;\;\;}l}
  V(P) - V(Q) & V(c) - V(c_1) - V(c_2) \\[5pt]
                & {{ \mid c_1 \mid \ \mid c_2 \mid } \over
              { \mid c_1 \mid + \mid c_2 \mid }}
              \Vert {\bf c_1} - {\bf c_2} \Vert ^2 \ 
        ,
\end{array}
\end{displaymath}
which is the dissimilarity given in Table \ref{tabhier}.  
This is a dissimilarity
which may be determined for any pair of classes of partition $P$; and the
agglomerands are those classes, $c_1$ and $c_2$, for which it is minimum.

It may be noted that if $c_1$ and $c_2$ are singleton classes, then 
$V(\{c_1,c_2\}) =
{1 \over 2} \Vert {\bf c}_1 - {\bf c}_2 \Vert^2 $, i.e.\ the variance
of a pair of objects is equal to half their Euclidean distance.

\section{Efficient Hierarchical Clustering Algorithms Using 
Nearest Neighbor Chains}
\label{sect2}

Early, efficient algorithms for hierarchical clustering are due to 
Sibson (1973), Rohlf (1973) and Defays (1977).
Their $O(n^2)$ implementations of the single
link method and of a (non-unique) complete link method, respectively, have
been widely cited.  

In the early 1980s a range of significant improvements 
(de Rham, 1980; Juan, 1982) were made 
to the Lance-Williams, or related, dissimilarity update schema, which 
had been in wide use since the mid-1960s.  Murtagh (1983, 1985) 
presents a 
survey of these algorithmic improvements.  We will briefly describe them here.
The new algorithms, which have the potential for {\em exactly} replicating
results found in the classical but more computationally expensive way,
are based on the construction of 
{\em nearest neighbor chains} and {\em reciprocal} or mutual NNs (NN-chains
and RNNs).

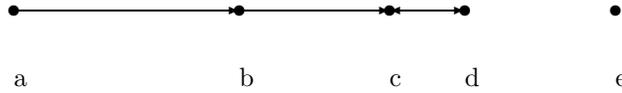
\begin{figure}[tb]
\begin{center}
\begin{minipage}[t]{5.0cm}
\setlength{\unitlength}{1cm}
\begin{pspicture}(0,3)(6,2.5)
\put(7,4){\circle*{0.15}}
\put(5,4){\circle*{0.15}}
\put(4,4){\circle*{0.15}}
\put(2,4){\circle*{0.15}}
\put(-1,4){\circle*{0.15}}
\put(7,3){e}
\put(5,3){d}
\put(4,3){c}
\put(2,3){b}
\put(-1,3){a}
\put(-1,4){\vector(1,0){3}}
\put(2,4){\vector(1,0){2}}
\put(4,4){\vector(1,0){1}}
\put(5,4){\vector(-1,0){1}}
\end{pspicture}
\end{minipage}
\end{center}
 \caption{Five points, showing NNs and RNNs.}
\label{figrnns}
\end{figure}

A NN-chain consists of an arbitrary point ($a$ in Fig.\ \ref{figrnns}); followed
by its NN ($b$ in Fig.\ \ref{figrnns}); followed by the NN from among the 
remaining points ($c$, $d$, and $e$ in Fig.\ \ref{figrnns}) of this 
second point;
and so on until we necessarily have some pair of points which can
be termed reciprocal or mutual NNs.  (Such a pair of RNNs may be
the first two points in the chain; and we have assumed that no
two dissimilarities are equal.)  

\begin{figure}[tb]
\centerline{\psfig{figure=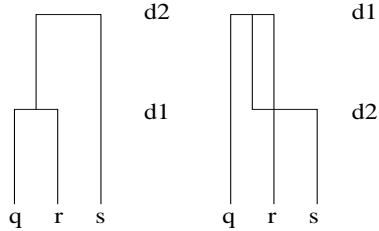,height=5cm,width=3cm,angle=270}}
  \caption{Alternative representations of a hierarchy with an inversion.
Assuming dissimilarities, as we go vertically up, agglomerative 
criterion values ($d_1$,
$d_2$) increase so that $d_2 > d_1$.  But here, undesirably, $d_2 < d_1$ 
and the ``cross-over'' or inversion (right panel) arises.}
\label{figinv}
\end{figure}

In constructing a NN-chain, irrespective of the starting point, 
we may agglomerate a pair of RNNs as soon as they are found.
What guarantees that we can arrive at the same hierarchy as if
we used traditional ``stored dissimilarities'' or ``stored data'' 
algorithms?  Essentially this is the
same condition as that under which no inversions or reversals
are produced by the clustering method. Fig.\ \ref{figinv} gives an
example of this, where $s$ is agglomerated at a lower criterion
value (i.e. dissimilarity) than was the case at the previous
agglomeration between $q$ and $r$.  Our ambient space has thus
contracted because of the agglomeration.  This is due to the algorithm
used -- in particular the agglomeration criterion -- and it is something
we would normally wish to avoid.

This is formulated as:

$$   \mbox{ Inversion impossible if: } \ 
      d(i,j) < d(i,k) \  {\rm or} \   d(j,k) $$
$$       \Rightarrow
      d(i,j) < d(i \cup j,k)$$

This is one form of Bruynooghe's {\em reducibility property} (Bruynooghe,
1977; see also Murtagh, 1984). 
Using the Lance-Williams dissimilarity update formula, it can
be shown that the minimum variance method does not give rise
to inversions; neither do the linkage methods; but the median
and centroid methods cannot be guaranteed {\em not} to have inversions.
    
To return to Fig.\ \ref{figrnns}, if we are dealing with a clustering
criterion which precludes inversions, then  $c$ and $d$
can justifiably be agglomerated, since no other point (for
example, $b$ or $e$) could have been agglomerated to either of these.

The processing required, following an agglomeration, is to
update the NNs of points such as $b$ in Fig.\ \ref{figrnns} (and on account
of such points, this algorithm was dubbed {\em algorithme
des c\'elibataires}, or bachelors' algorithm,
in de Rham, 1980).  The following is a
summary of the algorithm:

\vspace{.25in}

\noindent
{\bf NN-chain algorithm}
\begin{description}
\item[Step 1] Select a point arbitrarily.

\item[Step 2] Grow the NN-chain from this point until a pair of 
        RNNs  is obtained.

\item[Step 3] Agglomerate these points (replacing with a cluster
        point, or updating the dissimilarity matrix).

\item[Step 4] From the point which preceded the RNNs (or from any
        other arbitrary point if the first two points chosen
        in steps 1 and 2 constituted a pair of RNNs), return
        to step 2 until only one point remains.
\end{description}

In Murtagh (1983, 1984, 1985) and Day and Edelsbrunner (1984), one finds
discussions of $O(n^2)$ time and $O(n)$ space implementations of Ward's
minimum variance (or error sum of squares) method and of the centroid
and median methods.  The latter two methods are termed the UPGMC and WPGMC
criteria by Sneath and Sokal (1973).  Now, a problem with the cluster
criteria used by these latter two methods is that the reducibility property
is not satisfied by them.  This means that the hierarchy constructed may
not be unique as a result of inversions or reversals (non-monotonic variation)
in the clustering criterion value determined in the sequence of 
agglomerations.

Murtagh (1983, 1985) describes $O(n^2)$ time and $O(n^2)$ space implementations
for the single link method, the complete link method and for the weighted
and unweighted group average methods (WPGMA and UPGMA).  This approach is
quite general vis \`a vis the dissimilarity used and can also be used 
for hierarchical clustering methods other than those mentioned.  

Day and Edelsbrunner (1984) prove the exact $O(n^2)$ time complexity of 
the centroid and median methods using an argument related to the 
combinatorial problem of optimally packing hyperspheres into an 
$m$-dimensional volume.  They also address the question of metrics: 
results are valid in a wide class of distances including those associated
with the Minkowski metrics.  

The construction and maintenance of the nearest neighbor chain as well 
as the carrying out of agglomerations whenever reciprocal nearest neighbors
meet, both offer possibilities for distributed implementation.  
Implementations on a 
parallel machine architecture were described by Willett (1989).  

Evidently (from Table \ref{tabhier})  
both coordinate data and graph (e.g., dissimilarity) data can be
input to these agglomerative methods.  Gillet et al.\ (1998) in the 
context of clustering chemical structure databases refer to the common 
use of the Ward method, based on the reciprocal nearest neighbors algorithm,
on data sets of a few hundred thousand molecules.  

Applications of hierarchical clustering to bibliographic information 
retrieval are assessed in Griffiths et al.\ (1984).  Ward's minimum variance
criterion is favored.

From details in White and McCain (1997), the Institute of Scientific 
Information (ISI) clusters citations (science, and social science) by first 
clustering highly cited documents based on a single linkage criterion, 
and then four more passes are made 
through the data to create a subset of a single linkage hierarchical
clustering.

In the CLUSTAN and R statistical data analysis packages (in 
addition to 
hclust in R, see flashClust due to P.\ Langfelder and available 
on CRAN, ``Comprehensive R Archive Network'', cran.r-project.org) there are 
implementations of the NN-chain algorithm for the minimum variance 
agglomerative criterion.  
A property of the minimum variance agglomerative hierarchical 
clustering method is that we can use
weights on the objects on which we will induce a hierarchy.  By 
default, these weights are identical and equal to 1.   Such weighting 
of observations to be clustered is an important and practical aspect 
of these software packages.

\section{Hierarchical Self-Organizing Maps and Hierarchical Mixture 
Modeling}
\label{sect6}

It is quite impressive how 2D (2-dimensional or, for that matter, 3D) 
image signals can handle with ease the 
scalability limitations of clustering and many other data processing 
operations.  The contiguity imposed on adjacent pixels or grid cells
bypasses the 
need for nearest neighbor finding.  It is very interesting  
therefore to consider the feasibility of taking problems 
of clustering massive data sets into the 2D image domain.  
The Kohonen self-organizing feature map exemplifes this well.
In its basic variant (Kohonen, 1984, 2001) is can be formulated in terms of
k-means clustering subject to a set of interrelationships between
the cluster centers (Murtagh and Fern\'andez-Pajares, 1995).  

Kohonen maps lend themselves well for hierarchical representation.
Lampinen and Oja (1992), Dittenbach et al.\ (2002) and Endo et al.\ 
(2002) elaborate on the 
Kohonen map in this way.   An example application in character 
recognition is Miikkulanien (1990).  

A short, informative review of hierarchical self-organizing maps is provided
by Vicente and Vellido (2004).   
These authors also review what they term as probabilistic hierarchical 
models.  This includes putting into a hierarchical framework the 
following: Gaussian mixture models, and a probabilistic -- Bayesian --
alternative to the Kohonen self-organizing map termed Generative 
Topographic Mapping (GTM).  

GTM can be traced to the Kohonen self-organizing map in the following 
way.  Firstly, we consider the hierarchical map as brought about through
a growing process, i.e.\ the target map is allowed to grow in terms of
layers, and of grid points within those layers.   Secondly, we impose
an explicit probability density model on the data.   Tino and 
Nabney (2002) discuss how the local hierarchical models are organized 
in a hierarchical way.   

In Wang et al.\ (2000) an alternating Gaussian mixture modeling, and 
principal component analysis, is described, in this way furnishing a 
hierarchy of model-based clusters.  AIC, 
the Akaike information criterion, is used for selection of the best 
cluster model overall.

Murtagh et al.\ (2005) use a top level Gaussian mixture modeling with 
the (spatially aware) PLIC, pseudo-likelihood information criterion, 
used for cluster selection and identifiability.   Then at the next level
-- and potentially also for further divisive, hierarchical levels -- 
the Gaussian mixture modeling is continued but now using the marginal 
distributions within each cluster, and using the analogous Bayesian
clustering identifiability criterion which is the Bayesian information 
criterion, BIC.  The resulting output is referred to as a model-based
cluster tree.  

The model-based cluster tree algorithm of Murtagh et al.\ (2005) is a
divisive hierarchical algorithm.  Earlier in this article, we considered
agglomerative algorithms.  However it is often feasible to implement
a divisive algorithm instead, especially when a graph cut (for example) 
is important for the application concerned.   
Mirkin (1996, chapter 7) describes divisive Ward, minimum variance 
hierarchical clustering, which is closely related to a 
bisecting k-means also.  

A class of methods under the name of spectral clustering uses 
eigenvalue/eigenvector reduction on the (graph) adjacency matrix.  As 
von Luxburg (2007) points out in reviewing this field of spectral 
clustering, such methods have ``been discovered, re-discovered, 
and extended many times in different communities''.   Far from 
seeing this great deal of work on clustering in any sense in a 
pessimistic way, we see the perennial and pervasive interest in 
clustering  as testifying to the continual renewal and innovation 
in algorithm developments, faced with application needs.  

It is 
indeed interesting to note how the clusters in a hierarchical 
clustering may be {\em defined} by the eigenvectors of a dissimilarity
matrix, but subject to carrying out the eigenvector reduction in 
a particular algebraic structure, a semi-ring with additive 
and multiplicative operations 
given by ``min'' and ``max'', respectively (Gondran, 1976).  

In the next section, section \ref{sect7}, the themes of mapping, and of
divisive algorithm, are frequently taken in a somewhat different 
direction.   As always, the application at issue is highly relevant
for the choice of the hierarchical clustering algorithm.   

\section{Density and Grid-Based Clustering Techniques}
\label{sect7}

Many modern clustering techniques focus on large data sets.  In 
Xu and Wunsch (2008, p.\ 215) these are classified as follows:

\begin{itemize}
\item Random sampling
\item Data condensation
\item Density-based approaches
\item Grid-based approaches
\item Divide and conquer
\item Incremental learning
\end{itemize}

From the point of view of this article, we select 
density and grid based 
approaches, i.e., methods that either look for data 
densities or split the data space into cells when looking for 
groups. In this section we take a look at these 
two families of methods.

The main idea is to use a grid-like structure to split the 
information space, separating the dense grid regions from the 
less dense ones to form groups.

In general, a typical approach within this category will consist 
of the following steps as presented by Grabusts and Borisov (2002):
\begin{enumerate}
  \item Creating a grid structure, i.e.\ partitioning the data space 
into a finite number of  non-overlapping cells.
  \item Calculating the cell density for each cell.
  \item Sorting of the cells according to their densities.
  \item Identifying cluster centers.
  \item Traversal of neighbor cells.
\end{enumerate}

Some of the more important algorithms within this category are the following:

\begin{itemize}
  \item[--] \textbf{STING:} STatistical INformation Grid-based clustering was 
proposed by Wang et al.\ (1997) who divide the spatial area into 
rectangular cells represented by a hierarchical structure. The root is 
at hierarchical level 1, its children at level 2, and so on. This 
algorithm has a computational complexity of $O(K)$, where $K$ is the 
number of cells in the bottom layer.  This implies that scaling this 
method to higher dimensional spaces is difficult (Hinneburg and Keim, 
1999). For 
example, if in high dimensional data space each cell has four children, 
then the number of cells in the second level will be $2^{m}$, where $m$ 
is the dimensionality of the database.
    \item[--] \textbf{OptiGrid:} Optimal Grid-Clustering was introduced 
by Hinneburg and Keim (1999) as an efficient algorithm to cluster 
high-dimensional databases with noise. It uses data partitioning based 
on divisive recursion by multidimensional grids, focusing on separation 
of clusters by hyperplanes. A cutting plane is chosen which goes through 
the point of minimal density, therefore splitting two dense half-spaces.  
This process is applied recursively with each subset of data. This 
algorithm is hierarchical, with time complexity of $O(n \cdot m)$ 
(Gan et al., 2007, pp.\ 210--212).
      \item[--] \textbf{GRIDCLUS:} proposed by Schikute  (1996) is a 
hierarchical algorithm for clustering very large datasets. It uses a 
multidimensional data grid to organize the space surrounding the data 
values rather than organize the data themselves. Thereafter patterns 
are organized into blocks, which in turn are clustered by a topological 
neighbor search algorithm. Five main steps are involved in the GRIDCLUS 
method: (a) insertion of points into the grid structure, (b) calculation 
of density indices, (c) sorting the blocks with respect to their density 
indices, (d) identification of cluster centers, and (e) traversal of 
neighbor blocks.
        \item[--] \textbf{WaveCluster:} this clustering technique 
proposed by Sheikholeslami et al. (2000) defines a uniform two dimensional 
grid on the data and represents the data points in each cell by the 
number of points. Thus the data points become a set of grey-scale 
points, which is treated as an image.  Then the problem of looking 
for clusters is transformed into an image segmentation problem, 
where wavelets are used to take advantage of their multi-scaling 
and noise reduction properties. The basic algorithm is as follows: 
(a) create a data grid and assign each data object to a cell in the 
grid, (b) apply the wavelet transform to the data, (c) use the 
average sub-image to find connected clusters (i.e.\ connected pixels), 
and (d) map the resulting clusters back to the points in the original 
space.  There is a great deal of other work also that is based on 
using the wavelet and other multiresolution transforms for 
segmentation.  
\end{itemize}
Further grid-based clustering algorithms can be found in the following: 
Chang and Jin (2002), Park and Lee (2004), Gan et al.\ (2007), and Xu 
and Wunsch (2008).

Density-based clustering algorithms are defined as dense regions of 
points, which are separated by low-density regions.  Therefore, clusters
can have an arbitrary shape and the points in the clusters may be 
arbitrarily 
distributed. An important advantage of this methodology is that only one 
scan of the dataset is needed and it can handle noise effectively.  
Furthermore the number of clusters to initialize the algorithm is not
required.

Some of the more important algorithms in this category include the 
following:

\begin{itemize}
  \item[--] \textbf{DBSCAN:} Density-Based Spatial Clustering of 
Applications with Noise was proposed by Ester et al.\ (1996)
to discover arbitrarily shaped clusters. Since it finds clusters based
 on density it does not need to know the number of clusters at 
initialization time. This algorithm has been widely used and has  
many variations (e.g., see GDBSCAN by Sander et al.\ 
(1998), PDBSCAN by Xu 
et al.\ (1999), 
and DBCluC by Za\"{\i}ane and Lee (2002).
    \item[--] \textbf{BRIDGE:} proposed by Dash et al.\ (2001) uses a hybrid 
approach integrating  $k$-means to partition the dataset into $k$ 
clusters, and then density-based algorithm DBSCAN is applied to each 
partition to find dense clusters.
      \item[--] \textbf{DBCLASD:} Distribution-Based Clustering of LArge 
Spatial Databases (see Xu et al., 1998) 
assumes that data points within a cluster 
are uniformly distributed. The cluster produced is defined in terms of 
the nearest neighbor distance.
        \item[--] \textbf{DENCLUE:} DENsity based CLUstering aims to 
cluster large multimedia data. It can find arbitrarily shaped clusters 
and at the same time deals with noise in the data. This algorithm has 
two steps.  First a pre-cluster map is generated, and the data is divided in 
hypercubes where only the populated are considered. The second step takes 
the highly populated cubes and cubes that are connected to a highly 
populated cube to produce the clusters. For a detailed presentation of 
these steps see Hinneburg and Keim (1998).
          \item[--] \textbf{CUBN:} this has three steps. First an erosion 
operation is carried out to find border points. Second, the nearest 
neighbor method is used to cluster the border points. Finally, the 
nearest neighbor method is used to cluster the inner points. This algorithm 
is capable of finding non-spherical shapes and wide variations in size. 
Its computational complexity is $O(n)$ with $n$ being the size of the 
dataset. For a detailed presentation of this algorithm see Wang and 
Wang (2003).
\end{itemize}

\section{A New, Linear Time Grid Clustering Method: 
m-Adic Clustering}
\label{sect8}

In the last section, section \ref{sect7}, 
 we have seen a number of clustering methods that 
split the data space into cells, cubes, or dense regions to locate
high density areas that can be further studied to find clusters.

For large data sets clustering via an $m$-adic ($m$ integer, which if
a prime is usually denoted as $p$) expansion is possible,
with the advantage of doing so in linear time for the clustering
algorithm based on this expansion.  The usual
base 10 system for numbers is none other than the case of  $m = 10$ 
and the base 2 or binary system 
can be referred to as 2-adic where $p = 2$.  Let us 
consider the following distance relating to 
the case of vectors 
$x$ and $y$  with 1 attribute, hence unidimensional:

\begin{equation}
\label{eq:baire}
\db(x, y) =   
	\left\{ 
	\begin{array}{ll}
       1 &\;\; $if$\;\;  x_{1} \neq y_{1}\\
       $inf$\;\;  m^{-k} & \;\;\;\;\;\; x_{k} = y_{k} \;\;\; 1 \leq k \leq
       \left| K \right|
    \end{array}
    \right.
\label{eqnbaire}
\end{equation}

This distance defines the longest common prefix of strings.  A 
space of strings, with this distance, is a 
Baire space.  Thus we call this the Baire distance: here the longer the common 
prefix, the closer a pair of sequences. What is of interest to us here 
is this longest common prefix metric, which is an ultrametric (Murtagh et al., 
2008).

For example, let us consider two such values, $x$ and $y$.
We take $x$ and $y$ to be bounded by 0 and 1. Each are of some 
precision, and we take the integer $\left| K \right|$ to be the 
maximum precision.  

Thus we consider ordered sets $x_{k}$ and $y_{k}$ for $k \in K$. 
So, $k = 1$ is the index of 
the first decimal 
place of precision; $k = 2$ is the index of the second decimal 
place; . . . ; 
$k = \left| K \right|$ is the index of the 
$\left| K \right|$th decimal place.  
The cardinality of the set K is the precision with which a number, 
$x$, is measured.

Consider as examples $x_{k} = 0.478$; and $y_{k} = 0.472$. In these 
cases, $\left| K \right| = 3$. Start from the first decimal position.
For $k = 1$, we have $x_{k} = y_{k} = 4$. For $k = 2$, $x_{k} = y_{k}$. 
But for $k = 3$,  $x_{k} \neq y_{k}$.  Hence their Baire distance is 
$10^{-2}$ for base $m = 10$.  

It is seen that this distance splits a unidimensional string of 
decimal values
 into a 10-way
hierarchy, in which each leaf can be seen as a grid cell.  From 
equation (\ref{eqnbaire}) we can read off the distance 
between points assigned
to the same grid cell.  All pairwise distances of points assigned to 
 the same cell are the same.  

Clustering using this Baire distance has been successfully applied 
to areas such as 
chemoinformatics  (Murtagh et al., 2008), astronomy 
(Contreras and Murtagh, 2009) and text retrieval (Contreras, 2010).

\section{Conclusions}

Hierarchical clustering methods, with roots going back to the 
1960s and 1970s, are continually replenished with new challenges. 
As a family of algorithms they are central to the addressing of many 
important problems.  Their deployment in  many application domains
testifies to how hierarchical clustering methods will remain 
crucial for a long time to come.  

We have looked at both traditional agglomerative hierarchical 
clustering, and more recent developments in grid or cell based 
approaches.   We have discussed various algorithmic aspects, 
including well-definedness (e.g.\ inversions) and computational 
properties.   We have also touched on a number of application 
domains, again in areas that reach back over some decades 
(chemoinformatics) or many decades (information retrieval,
which motivated much early work in clustering, including 
hierarchical clustering), and 
more recent application domains (such as hierarchical model-based
clustering approaches). 

\section{References}

\begin{enumerate}
\item Anderberg MR {\em Cluster Analysis for Applications}.
       Academic Press, New York, 1973.


\item Benz\'ecri et coll., JP {\em L'Analyse des Donn\'ees. I. La
       Taxinomie}, Dunod, Paris, 1979 (3rd ed.).

\item Blashfield RK and Aldenderfer MS The literature on cluster
       analysis {\em Multivariate Behavioral Research} 1978, 
       13: 271--295.


\item Bruynooghe M M\'ethodes nouvelles en classification automatique
des donn\'ees taxinomiques nombreuses {\em Statistique et Analyse des
Donn\'ees} 1977,  no.\ 3, 24--42.

\item
Chang J-W and  Jin D-S, A new cell-based clustering method for large, 
high-dimensional data in data mining applications, in SAC '02: Proceedings 
of the 2002 ACM Symposium on Applied Computing. New York: ACM, 2002, 
pp. 503--507.

\item Contreras P Search and Retrieval in Massive Data Collections 
PhD Thesis. Royal Holloway, University of London, 2010.

\item Contreras P and Murtagh F Fast hierarchical clustering from the Baire 
distance.  In 
Classification as a Tool for Research, eds. H. Hocarek-Junge and 
C. Weihs, Springer, Berlin, 235--243, 2010.

\item
Dash M, Liu H, and  Xu X, $1 + 1 > 2$: Merging distance and density 
based clustering, in DASFAA '01: Proceedings of the 7th International 
Conference on Database Systems for Advanced Applications. Washington, 
DC: IEEE Computer Society, 2001, pp. 32--39.

\item Day WHE and Edelsbrunner H Efficient algorithms for 
agglomerative hierarchical clustering methods {\em Journal of Classification}
1984, 1: 7--24.

\item Defays D An efficient algorithm for a complete link method
{\em Computer Journal} 1977, 20: 364--366.

\item de Rham C La classification hi\'erarchique ascendante selon 
la m\'ethode des voisins r\'eciproques {\em Les Cahiers de l'Analyse des
Donn\'ees} 1980, V: 135--144.

\item Deza MM and Deza E {\em Encyclopedia of Distances}. 
Springer, Berlin, 2009.

\item 
Dittenbach M, Rauber A and  Merkl D Uncovering the hierarchical
structure in data using the growing hierarchical self-organizing map
{\em Neurocomputing}, 2002, 48(1--4):199--216.

\item 
Endo M, Ueno M and and Tanabe T
A clustering method using hierarchical self-organizing maps {\em 
Journal of VLSI Signal Processing} 32:105--118, 2002.  

\item
Ester M,  Kriegel H-P, Sander J, and Xu  X, A density-based algorithm 
for discovering clusters in large spatial databases with noise, in 2nd 
International Conference on Knowledge Discovery and Data Mining. 
AAAI Press, 1996, pp. 226--231.

\item
Gan G, Ma C and  Wu J 
{\em Data Clustering Theory, Algorithms, and Applications}  
Society for Industrial and Applied Mathematics. SIAM, 2007.

\item Gillet VJ, Wild DJ, Willett P and Bradshaw J 
Similarity and dissimilarity methods for processing chemical structure
databases {\em Computer Journal} 1998, 41: 547--558.

\item Gondran M Valeurs propres et vecteurs propres en 
classification hi\'erarchique {\em RAIRO Informatique Th\'eorique}
1976, 10(3): 39--46. 

\item Gordon AD {\em Classification}, Chapman and Hall, London, 1981.

\item Gordon AD A review of hierarchical classification
{\em Journal of the Royal Statistical Society A} 1987, 150: 119--137.

\item
Grabusts P and Borisov A Using grid-clustering methods in data classification,
in PARELEC '02: Proceedings of 
the International Conference on Parallel Computing in 
Electrical Engineering.Washington, DC: IEEE Computer Society, 2002.

\item Graham RH and Hell P On the history of the minimum spanning tree
problem {\em Annals of the History of Computing} 1985 7: 43--57.

\item Griffiths A, Robinson LA and Willett P 
Hierarchic
agglomerative clustering methods for automatic document classification 
{\em  Journal
of Documentation} 1984, 40: 175--205.

\item
Hinneburg  A and  Keim DA, 
A density-based algorithm for discovering clusters 
in large spatial databases with noise, in Proceeding of the 4th International 
Conference on Knowledge Discovery and Data Mining.  New York: AAAI 
Press, 1998, pp. 58--68.

\item
Hinneburg A and Keim D 
Optimal grid-clustering: Towards breaking the curse of 
dimensionality in high-dimensional clustering, in VLDB '99: Proceedings of the 
25th International Conference on Very Large Data Bases. San Francisco, CA: 
Morgan Kaufmann Publishers Inc., 1999, pp. 506--517.


\item 
Jain AK and Dubes RC {\em Algorithms For Clustering Data}
Prentice-Hall, Englwood Cliffs, 1988.         

\item Jain AK, Murty, MN and  Flynn PJ Data clustering: a review
{\em ACM Computing Surveys} 1999, 31: 264--323.

\item Janowitz, MF {\em Ordinal and Relational Clustering}, 
World Scientific, Singapore, 2010.  

\item Juan J Programme de classification hi\'erarchique par 
l'algorithme de la recherche en cha\^{\i}ne des voisins r\'eciproques
{\em Les Cahiers de l'Analyse des Donn\'ees} 1982, VII: 219--225.

\item 
Kohonen T {\em Self-Organization and Associative Memory} Springer, Berlin,
1984.

\item
Kohonen T {\em Self-Organizing Maps}, 3rd edn., Springer, Berlin, 2001.

\item 
Lampinen J and Oja E Clustering properties of hierarchical
self-organizing maps {\em Journal of Mathematical Imaging and Vision} 
2: 261--272, 1992.

\item 
Lerman, IC {\em Classification et Analyse Ordinale des
Donn\'ees}, Dunod, Paris, 1981.  

\item 
Le Roux B and Rouanet H {\em Geometric Data Analysis: From 
Correspondence Analysis to Structured Data Analysis}, Kluwer, 
Dordrecht, 2004. 

\item von Luxburg U A tutorial on spectral clustering {\em 
Statistics and Computing} 1997, 17(4): 395--416. 

\item March ST Techniques for structuring database records
{\em ACM Computing Surveys} 1983, 15: 45--79.

\item 
Miikkulainien R Script recognition with hierarchical feature maps
{\em Connection Science} 1990, 2: 83--101.


\item Mirkin B {\em Mathematical Classification and Clustering} 
Kluwer, Dordrecht, 1996. 

\item Murtagh F A survey of recent advances in hierarchical clustering
algorithms {\em Computer Journal} 1983, 26, 354--359.

\item Murtagh F Complexities of hierarchic clustering algorithms: 
state of the art {\em Computational Statistics Quarterly} 1984, 1: 101--113.



\item Murtagh F {\em Multidimensional Clustering Algorithms}
Physica-Verlag, W\"urzburg, 1985.


\item
Murtagh F and Hern\'andez-Pajares M 
The Kohonen self-organizing map method: an assessment, {\em Journal of 
Classification} 1995 12, 165-190.

%



\item Murtagh F, Raftery AE and Starck JL Bayesian inference for 
multiband image segmentation via model-based clustering trees, 
{\em Image and Vision Computing} 2005, 23: 587--596.

\item Murtagh F 
{\em Correspondence Analysis and Data Coding with Java and R},
Chapman and Hall, Boca Raton, 2005.

\item Murtagh F The Haar wavelet transform of a dendrogram
{\em Journal of Classification} 2007, 24: 3--32.


\item 
Murtagh F Symmetry in data mining and analysis: a unifying view 
based on hierarchy {\em Proceedings of Steklov Institute of Mathematics}
2009,  265: 177--198.




\item
Murtagh F and Downs G and Contreras P  Hierarchical clustering of massive, 
high dimensional data sets by exploiting ultrametric embedding
{\em SIAM Journal on Scientific Computing} 2008, 30(2): 707--730.

\item
Park NH and Lee WS, Statistical grid-based clustering over data streams,
{\em SIGMOD Record} 2004, 33(1): 32--37.

 \item Rapoport A and Fillenbaum S An experimental study of semantic
       structures, in Eds. A.K. Romney, R.N. Shepard and S.B. Nerlove,
       {\em Multidimensional Scaling; Theory and Applications in the
       Behavioral Sciences. Vol. 2, Applications}, Seminar Press, New York,
       1972, 93--131.

\item Rohlf FJ Algorithm 76: Hierarchical clustering using the 
minimum spanning tree {\em Computer Journal} 1973,  16: 93--95.

\item
Sander J, Ester M, Kriegel H.-P, and  Xu X, Density-based clustering in 
spatial databases: The algorithm GDBSCAN and its applications
{\em Data Mining 
Knowledge Discovery} 1998, 2(2): 169--194.

\item
Schikuta E Grid-clustering: An efficient hierarchical clustering method for 
very large data sets, in ICPR '96: Proceedings of the 13th International Conference 
on Pattern Recognition. Washington, DC: IEEE Computer Society, 1996, pp. 
101--105.

\item
Sheikholeslami G,  Chatterjee S and  Zhang A, Wavecluster: a wavelet based 
clustering approach for spatial data in very large databases, 
{\em The VLDB Journal}, 2000, 8(3--4): 289--304.

\item Sibson R  SLINK: an optimally efficient algorithm for the 
single link cluster method {\em Computer Journal} 1973, 16: 30--34.

 \item Sneath PHA and Sokal RR {\em Numerical Taxonomy}, Freeman,
       San Francisco, 1973.

\item Tino P and Nabney I Hierarchical GTM: constructing localized 
non-linear projection manifolds in a principled way, 
{\em IEEE Transactions on Pattern Analysis and Machine Intelligence}, 
2002, 24(5): 639--656.

\item 
van Rijsbergen CJ {\em Information Retrieval} Butterworths, 
London, 1979 (2nd ed.).


\item
Wang L and Wang Z-O, CUBN: a clustering 
algorithm based on density and distance,
in Proceeding of the 2003 International Conference on Machine Learning and Cybernetics.
IEEE Press, 2003, pp. 108--112.

\item
Wang W, Yang J and Muntz R STING: A statistical information grid approach to 
spatial data mining, in VLDB '97: Proceedings of the 23rd International Conference 
on Very Large Data Bases.San Francisco, CA: Morgan Kaufmann Publishers 
Inc., 1997, pp. 18--195.

\item Wang Y, Freedman M.I. and Kung S.-Y. Probabilistic principal 
component subspaces: A hierarchical finite mixture model for data 
visualization, 
{\em IEEE Transactions on Neural Networks} 2000, 11(3): 625--636.

\item White HD and McCain KW Visualization of literatures,
in M.E. Williams, Ed., {\em Annual Review of Information Science and 
Technology (ARIST)} 1997, 32: 99--168.

\item Vicente D and Vellido A review of hierarchical models for data 
clustering and visualization. In R. Gir\'aldez, J.C. Riquelme and J.S. 
Aguilar-Ruiz, Eds., Tendencias de la Miner{\'i}a de Datos en Espa{\~n}a. 
Red Espa{\~n}ola de Miner\'ia de Datos, 2004.

\item Willett P Efficiency of hierarchic agglomerative clustering 
using the ICL distributed array processor {\em Journal of Documentation}
1989, 45: 1--45.

 \item Wishart D Mode analysis: a generalization of nearest neighbour which
       reduces chaining effects, in Ed. A.J. Cole, {\em Numerical
       Taxonomy}, Academic Press, New York, 282--311, 1969.

\item 
Xu R and Wunsch D Survey of clustering algorithms {\em IEEE 
Transactions on Neural Networks} 2005, 16: 645--678. 

\item 
Xu R and Wunsch DC {\em Clustering} IEEE Computer Society Press, 2008.

\item 
Xu X,  Ester M, Kriegel H-P and Sander J A distribution-based clustering 
algorithm for mining in large spatial databases, in ICDE '98: Proceedings 
of the Fourteenth International Conference on Data Engineering.
Washington, DC: IEEE Computer Society, 1998, pp.\ 324--331.

\item
Xu X, J\"{a}ger J, and Kriegel, H-P A fast parallel clustering algorithm 
for large spatial databases, {\em Data Mining Knowledge Discovery} 
1999, 3(3): 263--290.

\item
Za\"{\i}ane OR  and  Lee C-H, Clustering spatial 
data in the presence of obstacles: 
a density-based approach, in IDEAS '02: Proceedings of the 2002 International 
Symposium on Database Engineering and Applications.Washington, DC: 
IEEE Computer Society, 2002, pp. 214--223.

\end{enumerate}

\end{document}